\documentclass{proclund}

\newcommand{\Eg}{\mbox{E$_\gamma$}}
\newcommand{\Ogpp}{\mbox{$^{16}$O($\gamma$,pp)}}
\newcommand{\Ognp}{\mbox{$^{16}$O($\gamma$,np)}}
\newcommand{\Oeepp}{\mbox{$^{16}$O(e,e$'$pp)}}

\newcommand{\Cfour}{\mbox{$^{14}$C}}

\newcommand{\gpp}{\mbox{($\gamma$,pp)}}
\newcommand{\eepp}{\mbox{(e,e$'$pp)}}

\newcommand{\eep}{\mbox{(e,e$'$p)}}
\def\bra#1{\left\langle #1\right|}
\def\ket#1{\left| #1\right\rangle}
\begin{document}
\pagestyle{prochead}


\title{RELEVANCE OF HIGH-MOMENTUM NUCLEONS FOR NUCLEAR PHENOMENA}
\author{W. H. Dickhoff}
  \email{wimd@wuphys.wustl.edu}
  \homepage{http://www.physics.wustl.edu/~wimd}
\affiliation
 {Laboratory of Theoretical Physics, University of Gent,
 Proeftuinstraat 86, B-9000 Gent, Belgium\\~\\
}
\affiliation
 {Department of Physics, Washington University,
 St. Louis, Missouri 63130, USA\\~\\
}
\author{E. P. Roth}
\affiliation
 {Department of Physics, Washington University,
 St. Louis, Missouri 63130, USA\\~\\
}
\author{M. Radici}
\affiliation
 {Istituto Nazionale di Fisica Nucleare, Sezione di Pavia, Pavia, Italy,
 I-27100\\~\\
}

\begin{abstract}
  A brief review is given concerning the status of the theoretical work on
  nucleon spectral functions. A recent concern about the validity of the 
  concept of spectroscopic factors as deduced from \eep\ reactions at higher
  $Q^2$, is discussed in some detail. The consequences of the observed spectral
  strength are then considered in the context of nuclear saturation. It is 
  argued that
  short-range correlations are mainly responsible for the actual value of 
  the observed charge density in ${}^{208}\rm Pb$ and by extension for the 
  empirical value of the saturation density of nuclear matter.  
  This observation combined with the general understanding of the spectroscopic
  strength suggests that a renewed study of nuclear matter, emphasizing the
  self-consistent determination of the spectral strength due to short-range
  and tensor correlations, may shed light on the perennial
  nuclear saturation problem.
  First results using such a scheme are presented.
\end{abstract}
\maketitle
\setcounter{page}{1}

\section{Introduction}

During the last fifteen years considerable progress has been made in 
elucidating the limits of the nuclear mean-field picture.
The primary tool in exhibiting these limits in a quantitative fashion
has been provided by the \eep\ reaction~\cite{diehu,sihu,lap,pasihu}. 
In this paper the status of the theoretical understanding of the spectroscopic 
factors that have been deduced from the analysis of this reaction will be
reviewed briefly.
The qualitative features of the strength distribution can be understood by 
realizing that a considerable mixing occurs between hole states and two-hole 
one-particle (2h1p) states. This leads to the observed fragmentation pattern
which exhibits a single peak for valence hole states near the Fermi energy,
albeit with reduced strength. A strongly fragmented strength 
distribution is observed for more deeply bound states.
This fragmentation is due
to the strong coupling to 2h1p states and the presence of these states at
energies corresponding to these more deeply bound hole states which leads to
correspondingly small energy denominators.
For quantitative results one also 
requires the inclusion of short-range and tensor
correlations. On the one hand, this leads to a global depletion of mean-field
orbitals which ranges from 10\% in light nuclei to about 15\% in heavy nuclei
and nuclear matter~\cite{dic1,dic2}.
This depletion effect, on the other hand,
is then compensated by the admixture of high-momentum
components in the ground state.
These high-momentum nucleons have not yet been unambiguously identified 
experimentally using the \eep\ reaction. The search for these high-momentum
components in valence states has not been successful~\cite{bob,blom}, as
was anticipated by earlier theoretical work~\cite{mudi94}. 

A recent publication~\cite{louk12C} has challenged the conventional
interpretation of the \eep\ reaction with regard to valence hole states.
This challenge consists in questioning the validity of the constancy of the 
spectroscopic factor as a function of the four-momentum, $Q^2$, transferred by 
the virtual photon to the knocked-out nucleon.
This recent work employs a Skyrme-Hartree-Fock bound-state wave functions for
the initial proton, a Glauber-type description of the final-state interaction 
of the outgoing proton, and a factorization approximation for the 
electromagnetic vertex for the description of the reaction at higher $Q^2$.
The results obtained in Ref.~\cite{louk12C} display an
increasing spectroscopic strength with increasing $Q^2$ for the ${}^{12}{\rm
C}$ nucleus.
While the theoretical definition of the spectroscopic factor is unambiguously
independent of the probe, it is worth studying the description of the data at 
higher $Q^2$ in a consistent manner. To this end we employ a recently 
developed eikonal approximation~\cite{rada,radb,radc,rade,radd}
to describe the outgoing proton under 
conditions appropriate for a recent JLab experiment~\cite{16OJLab}. 
This description of the final-state-interaction (FSI) is combined with
previous results for the quasihole wave functions obtained for
${}^{16}$O~\cite{mupodi} which were employed for the description~\cite{rapo}
of a low $Q^2$ experiment~\cite{leus}. 
The absorption of the outgoing proton in the eikonal approach is then
related to the corresponding absorption experienced by a nucleon in nuclear
matter as obtained from the self-energy.
This self-energy is obtained from self-consistent calculations of the
nucleon spectral functions including the effect of short-range and tensor 
correlations in nuclear matter~\cite{libth}.
This approach for the analysis at $Q^2 = 0.8 {\rm (GeV/c)^2}$ is discussed
in this contribution and some initial results are presented.

Based on the experimental and theoretical results for the spectral strength
distribution, one may wonder what the consequences are for the ``energy'' or
``Koltun'' sum rule~\cite{miga,kolt,moug}.
In principle, one can ascertain that a perfect agreement of the theoretical
strength with the experimental one, will yield a correspondingly good agreement
for the energy per particle provided three-body forces are not too important.
Initial indications of the relevance of high-momentum nucleons, which sofar
have not been observed directly, for the question of the energy per 
particle have already been raised in earlier work~\cite{mupodi}.
In addition, we will argue in this work that the actual value of the
nuclear sturation density is dominated by the effects of short-range and tensor
correlations (SRC).
Recent experimental work will be discussed which supports this
claim.
Based on these considerations, it is sugested that a renewed study of the
nuclear satuation problem is in order.
Results will be discussed which suggests that new insights may be
obtained using this approach. We close with some conclusions.

\section{Status of Theoretical Results for Spectroscopic Strength}

One of the critical experimental ingredients in clarifying the nature
of nuclear correlations has only become available over the last decade
and a half. It is therefore not surprising that all schemes that
have been developed to calculate nuclear matter saturation properties
are not based on the insights that these experiments provide.
Before discussing the implications of these insights, we will review
these results in this section.
Exclusive experiments, involving the removal of a proton from the nucleus
which is induced by a high-energy electron that
is detected in coincidence with the
removed proton, have given access to absolute spectroscopic factors
associated with quasihole states for a wide range of
nuclei.~\cite{diehu,sihu,lap,pasihu}
The experimental results indicate that the removal of single-particle
strength for quasihole states
near the Fermi energy corresponds to about 65\%.
The spectroscopic factors obtained in these experiments can be directly related
to the single-particle Green's function of the system which is given by
\begin{eqnarray}
g(\alpha , \beta ; \omega ) & = &
\sum_m\ \frac{\bra{\Psi_0^A} a_\alpha \ket{\Psi_m^{A+1}}
\bra{\Psi_m^{A+1}} a_\beta^\dagger \ket{\Psi_0^A}}
{\omega - (E_m^{A+1} - E_0^A ) + i\eta} \nonumber \\
& + & \sum_n\ \frac{\bra{\Psi_0^A} a_\beta^\dagger \ket{\Psi_n^{A-1}}
\bra{\Psi_n^{A-1}} a_\alpha \ket{\Psi_0^A}}
{\omega - (E_0^A - E_n^{A-1} ) - i\eta} .
\label{eq:prop}
\end{eqnarray}
This representation of the Green's function is referred to as the 
Lehmann-representation and involves the exact eigenstates 
and corresponding energies of the
$A$- and $A\pm 1$-particle systems.
Both the addition and removal amplitude for a particle from (to)
the ground state of the system with $A$ particles must be considered
in Eq.~(\ref{eq:prop}).
Only the removal amplitude has direct relevance for the analysis
of the \eep\ experiments.
The spectroscopic factor for the removal of a particle in the single-particle
orbit $\alpha$, while leaving the remaining nucleus in state $n$,
is then given by
\begin{equation}
z_\alpha = \left| \bra{\Psi_n^{A-1}} a_\alpha \ket{\Psi_0^A} \right|^2 ,
\label{eq:specf}
\end{equation}
which corresponds to the contribution to the numerator of the second sum
in Eq.~(\ref{eq:prop}) of state $n$ for the case $\beta = \alpha$.
Another important quantity, which also contains this information,
is the spectral
function associated with single-particle orbit $\alpha$.
The part corresponding to the removal of particles, or hole spectral function,
is given by
\begin{equation}
S_h(\alpha , \omega) = \sum_n\
\left| \bra{\Psi_n^{A-1}} a_\alpha \ket{\Psi_0^A} \right|^2 
\delta(\omega - (E_0^A - E_n^{A-1} ) ) ,
\label{eq:spefu}
\end{equation}
which corresponds to the imaginary part of the diagonal elements
of the propagator and characterizes the strength distribution of
the single-particle state $\alpha$ as a function of energy in the
$A-1$-particle system.
From this quantity one can therefore obtain
another key ingredient that gauges the
effect of correlations, namely the occupation number which is given by
\begin{equation}
n(\alpha) = \int_{-\infty}^{\epsilon_F} d\omega\ S_h(\alpha , \omega )
= \bra{\Psi_0^A} a_\alpha^\dagger a_\alpha \ket{\Psi_0^A} .
\label{eq:occ}
\end{equation}
In the experimental analysis the quantum number $\alpha$ is related to the
actual Woods-Saxon potential required to both reproduce the correct energy of
the hole state as well as the shape of the corresponding
\eep\ cross section for this particular transition. 
The remaining parameter required to fit the actual data then 
becomes the spectroscopic factor associated with this transistion.
In this analysis the reduction of the flux associated with the
scattering of the outgoing proton is incorporated by the use of empirical
optical potentials describing elastic proton-nucleus scattering data. 
Experiments on ${}^{208}{\rm Pb}$ result in a spectroscopic factor of 0.65
for the removal of the last $3s_{1/2}$ proton~\cite{sihu}.
Additional information about the occupation number of this orbit can be 
obtained by analyzing elastic electron scattering cross sections of 
neighboring nuclei~\cite{wag}.
The actual occupation number for the $3s_{1/2}$ proton orbit obtained
from this analysis is about 10\% larger than the quasihole spectroscopic
factor~\cite{grab,grabx}.
A recent analysis of the \eep\ reaction on ${}^{208}{\rm Pb}$
in a wide range of missing energies and for missing momenta below 270 MeV/c
yields information on the occupation numbers of more deeply bound orbitals.
The data suggest that all deeply bound orbits are depleted by the same
amount of about 15\%~\cite{lapx,bat}.

As discussed in the introduction, the general properties of the 
experimental strength distributions can be understood on the basis of
the coupling between single-hole states and 2h1p states.
This implies that a proper inclusion of this coupling in the low-energy
domain is required in theoretical calculations that aim at reproducing the
experimental distribution of the strength.
Such calculations have been successfully performed for medium-heavy 
nuclei~\cite{brand,rijs}.
Indeed, calculations for the strength distribution for the removal of protons
from ${}^{48}{\rm Ca}$ demonstrate that an excellent qualitative agreement
with the experimental results is obtained when the coupling of the single-hole
states to low-lying collective states is taken into account~\cite{rijs}.
This coupling is taken into account by calculating the microscopic RPA
phonons and then constructing the corresponding self-energy.
The solution of the Dyson equation then provides the theoretical
strength distribution.
By adding the additional depletion due to short-range correlations a
quantitative agreement is obtained although no explicit calculation
for these nuclei including both effects has been performed to date.
The corresponding occupation numbers calculated for this nucleus
also indicate that the influence of collective low-lying states, associated
with long-range correlations, on the occupation numbers is confined to
single-particle states in the immediate vicinity of the
Fermi level.
The description of the spectroscopic strength in ${}^{16}$O is not as 
successful~\cite{geurts} on account of the complexity of the low-energy
structure of this nucleus.
Attempts to describe the proper inclusion of microscopic particle-particle
and particle-hole phonons in a Faddeev approach for this nucleus are currently
in progress~\cite{bardic,bardic1}.

For a global understanding of the strength distribution it is also
necessary to account for the appearance of single-particle strength
at high momenta as a direct reflection of the influence of short-range
correlations.
These high-momentum nucleons make up for the missing strength that has
been documented in \eep\ experiments.
Results for ${}^{16}{\rm O}$~\cite{mudi94,mupodi} corroborate the expected
occupation of high-momenta but put their presence at high missing
energy.
This can be understood in terms of the admixture of a high-momentum nucleon
requiring 2h1p states which can accomodate this momentum through momentum
conservation.
Since two-hole states combine to small total pair momenta, one necessarily
needs a high-momentum nucleon (of about equal and opposite value to the
component to be admixed) with corresponding high excitation energy.
As a result, one expects to find high-momentum components predominantly
at high missing energy.
Recent experiments at JLab attempt at a quantitative assessment of the
strength of these high-momentum nucleons~\cite{rohe}.

\section{Consistent Approach to the Analysis of Spectroscopic Strength at high
$Q^2$}

Recent work~\cite{louk12C} has challenged the conventional
interpretation of the \eep\ reaction with regard to valence hole states.
An analysis of all data for the ${}^{12}$C nucleus at low
$Q^2$ is obtained in this work using the standard analysis technique 
developed by the NIKHEF group.
For experiments at higher $Q^2$ a different approach was used.
For example, Skyrme-Hartree-Fock bound-state wave functions are employed for 
which no spectroscopic factor at NIKHEF kinematics is quoted.
To account for final-state interactions of the proton, which absorbs
the virtual photon, a Glauber-type description is employed with some slight
adjustments of the input nucleon-nucleon cross section to account for 
in-medium effects~\cite{zhal}.
As is usual in this approach,
a factorization approximation for the 
electromagnetic vertex is employed. It is unclear whether this represents
a serious approximation.
At low $Q^2$, it would be unacceptable. 
The results of Ref.~\cite{louk12C} exhibit an
increase in the spectroscopic strength with increasing $Q^2$ for
the ${}^{12}{\rm C}$ nucleus.

The theoretical definition of the spectroscopic factor (see 
Eq.~(\ref{eq:specf})) involves a matrix element of a particle removal
operator between the ground state and an appropriate state in the system
with one particle less.
Clearly, no dependence on $Q^2$ can be generated by such an expression.
The extraction of spectroscopic factors clearly involves a detailed model
describing the exit of a strongly interacting particle from the nucleus.
It is therefore of great interest to pursue the question whether one
can extend the analysis involving spectroscopic factors to kinematical
conditions involving higher $Q^2$.
We have therefore developed an approach to test this possibility which
relies on a recently developed eikonal model of the FSI~\cite{rada,radb,radc,
rade,radd}.
This approach will be used to describe a recent JLab experiment on
${}^{16}$O~\cite{16OJLab} at $Q^2 = 0.8$ ${\rm (GeV/c)^2}$.
We proceed by describing some of the key features of this approach before
discussing the other ingredients of this calculation.

The nuclear response in exclusive \eep\ process can be parametrized  as
a bilinear product of matrix elements of the different helicity components of
the nuclear current, which describe 
the transition from the initial to the final hadronic states.
In the projection-operator approach and within the framework of the
Distorted-Wave Impulse Approximation 
(DWIA)~\cite{libro}, it is possible to project out of the total Hilbert space
a suitable 
channel where the matrix elements are written in a one-body representation as
\begin{equation}
J^\lambda_{\alpha s's}(\vec p_f, \vec q) = \int d\vec r d\sigma 
e^{{\rm i} \vec q \cdot \vec r} 
\chi_{\vec p_f s'}^{\left( - \right) \, *} (\vec r,\sigma) 
\hat J_\lambda (\vec q, \vec r,  \sigma) 
\phi_{\alpha s} (\vec r,\sigma) \, ,
\label{eq:ampl}
\end{equation}
where $\vec q$ is the momentum carried by the virtual photon and
$\vec p_f, s'$ are the momentum and spin of the detected nucleon, leaving a
hole in the residual nucleus with 
collective quantum numbers $\alpha$ and spin $s$. If the detected nucleon is
moving fast, the scattering wave function $\chi$, with incoming-wave boundary
conditions, is usually 
approximated by an eikonal wave through the Glauber method~\cite{Glauber}.
However, since for a 
fastly moving object the nuclear density can be considered roughly constant
inside all the nuclear volume (except for a small portion on the surface),
the eikonal wave function can be 
further approximated in lowest order by a plane wave with complex momentum 
$\vec P_f = \vec p_f + {\rm i} \vec p_I$ and $\vec p_I \parallel \vec p_f$:
\begin{equation}
J^\lambda_{\alpha s's}(\vec p_f, \vec q) = \sum_{\tilde s} \int d\vec r
e^{{\rm i} \vec q \cdot \vec r} 
\left( e^{{\rm i} \vec p_f \cdot \vec r} e^{- \vec p_I \cdot \vec r}
\right)^* \delta_{s' \tilde s} 
\langle \tilde s | \hat J_\lambda (\vec q, \vec r) | s \rangle 
\phi_{\alpha s} (\vec r) \, .
\label{eq:eik}
\end{equation}
The scattering wave function now corresponds to a uniformly damped plane wave,
where the damping is driven by ${\rm Im} (\vec P_f) \equiv \vec p_I$.
This corresponds to solving the
Schr\"odinger equation with a complex potential for a particle travelling 
through homogeneous nuclear matter, i.e.
\begin{equation}
\left( {- \hbar^2 \over 2m} \nabla^2 + \hat{V} + {\rm i} \hat{W} \right)
\chi = E \chi \> , 
\label{eq:schro}
\end{equation}
or, equivalently,
\begin{equation}
\left( E - {\hat V} - {\rm i} {\hat W} \right) \chi = 
\displaystyle{ { { \vec P_f \cdot \vec P_f } \over 2m}} \chi  \  = \  \left( 
\displaystyle{ { { \vec p_f^2 - \vec p_I^2 } \over 2m } } + {\rm i} \  
\displaystyle{ { { \vec p_f \cdot \vec p_I } \over m } } \right) \chi \> , 
\label{eq:schroop}
\end{equation}
from which a natural relationship between $p_I$ and the absorptive part
$W$ of the potential 
is deduced. If the outgoing proton is sufficiently energetic,
i.e. $p_f \ge 1$ GeV/$c$, 
and comes from a bound state with a momentum below the Fermi surface,
this approximation has been shown to give reliable
results~\cite{rada,radb,radc} with a constant $p_I \propto 
W/p_f$. 

The technical advantage of this approximation is that the representation
of Eq. (\ref{eq:eik}) 
in momentum space becomes completely analytical, provided that the integral
is extended in the complex plane:
\begin{eqnarray}
J^\lambda_{\alpha s's}(\vec p_f, \vec p_I, \vec q) &= &\sum_{\tilde s}
\int d\vec P
\delta (\vec P_f - \vec P - \vec q ) \delta_{s' \tilde s} 
\langle \tilde s | \hat J_\lambda (\vec q, \vec P) | s \rangle 
\phi_{\alpha s} (\vec P) \nonumber \\
&= &\langle s' | \hat J_\lambda (\vec q, \vec P_f - \vec q) | s \rangle 
\phi_{\alpha s} (\vec P_f - \vec q) \, .
\label{eq:eik-p}
\end{eqnarray}
The conditions for an analytical extension are two: a suitable extension of
the definition of 
the $\delta $ distribution in terms of complex variables, which automatically
connects to the damped plane wave $e^{{\rm i} \vec P \cdot \vec r}$ as
in the usual case with real variables;
the integrand must be analytical and vanish asymptotically for
$|\vec P| \rightarrow \infty$.
Both conditions have been explored and verified in Ref.~\cite{rade,radd}. 

The results presented for the E89003 kinematics have been obtained
replacing $\vec p_I$ by the value predicted from nuclear-matter
calculations (discussed in the next section) at the considered $Q^2$.
This procedure corresponds to considering the wave function of the proton
at the momentum corresponding to the kinematical setting. This wave function
exhibits a damping due to the imaginary part of the self-energy.
The dominant feature descibing this damping is obtained by associating a 
complex pole to the propagator which leads to
the exponential damping of the wave function. This procedure
can be used to obtain
the complex part of the momentum~\cite{dick98} appropriate for this problem.
\begin{figure}[htb]
  \begin{center}
    \parbox[b]{0.5\linewidth}{
    \includegraphics[width=1.00\linewidth]{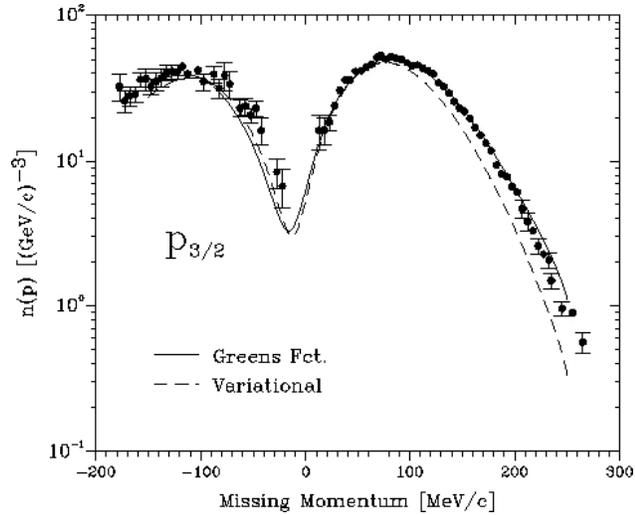}}
    \parbox[b]{0.7\linewidth}{
    \caption{\label{fig:spfux}
      Reduced cross section for the ${}^{16}$O\eep reaction in parallel 
      kinematics leading to the state at -6.32 MeV of the residual nucleus
      ${}^{15}$N. Results of Green's function calculations~\cite{mupodi}
      (solid line) are compared to those in the variational calculation of
      Ref.~\cite{piepra} (dashed line) and the experimental data~\cite{leus}.
      }}
    \end{center}
\end{figure}

\begin{figure}[htb]
  \begin{center}
    \parbox[b]{0.5\linewidth}{
      \includegraphics[width=0.8\linewidth]{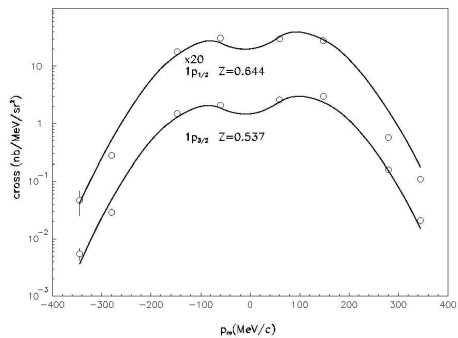}}
    \parbox[b]{0.35\linewidth}{
      \caption[experimental setup]{\label{fig:JLAB}
        Comparison of data obtained from the ${}^{16}$O\eep\ reaction at 
        $Q^2 = 0.8 {\rm (GeV/c)^2}$~\cite{16OJLab} with a consistent 
        theoretical approach involving the same spectroscopic factors
        as obtained for the NIKHEF experiment~\cite{leus}.}
      }
  \end{center}
\end{figure}

We then start this analysis by employing results for the quasihole wave
functions corresponding to $p_{3/2}$ (and $p_{1/2}$) 
protons from ${}^{16}$O obtained in 
Ref.~\cite{mupodi}.
These wave functions have been used to extract spectroscopic
factors~\cite{rapo} 
corresponding to a low $Q^2$ experiment performed at NIKHEF~\cite{leus}.
The result of this analysis is shown in Fig.~\ref{fig:spfux}.
The description of the data using the Green's function calculation of the
quasihole wave function require a spectroscopic factor of 0.537. With
this reduction factor a good description of the data is obtained.
Assuming the model for FSI described above and using the input from the
nucleon self-energy in nuclear matter to describe the damping of the outgoing
nucleon wave, we are in a position to calculate the cross section for the
JLab experiment at $Q^2 = 0.8$  ${\rm (GeV/c)^2}$
while keeping the {\bf same} spectroscopic
factors for the removal of the $p$-nucleons that generate a good description
of the data at low $Q^2$ as shown in Fig.~\ref{fig:spfux}.
The results of this calculation are shown in Fig.~\ref{fig:JLAB}.
These results suggest that it is indeed possible to use spectroscopic 
information obtained in the kinematical domain used at facilities like
NIKHEF and, subsequently,
correctly predict cross sections for the \eep\ reaction
at higher $Q^2$. We emphasize that no adjustment of any input of the
calculation has been performed to obtain the results in Fig.~\ref{fig:JLAB}.
Apparently, a consistent theoretical approach does not require a different
interpretation of the spectroscopic strength under different kinematical
conditions.

\section{Self-consistently Dressed Nucleons in Nuclear Matter} 

Having confirmed the reliability of the interpretation of the spectroscopic
strength obtained from the \eep\ reaction, it is possible to return to the
discussion of the consequences of these results. We first collect some relevant
information which will be used to argue that the empirical saturation density
of nuclear matter is dominated by SRC. 
As discussed earlier, a recent analysis of the \eep\ reaction on
${}^{208}{\rm Pb}$ up to 100 MeV missing energy and 270 MeV/c missing momenta
indicates that all deeply bound orbits are depleted by the same
amount of about 15\%~\cite{lapx,bat}.
This global depletion of the single-particle strength in about the same amount
for all deeply bound states, as observed for ${}^{208}{\rm Pb}$,
was anticipated~\cite{dimu,dic1} on the basis of the experience that has been
obtained
with calculating occupation numbers in nuclear matter with the inclusion
of SRC~\cite{von91}.
Such calculations suggest that about 15\% of the single-particle strength
in heavy nuclei is removed from the Fermi sea leading to the occupation of
high-momentum states. This global depletion of mean-field orbitals
can be interpreted as a clear signature of
the influence of SRC.
In turn, these results reflect on one of the key quantities determining
nuclear saturation empirically.
Elastic electron scattering from ${}^{208}{\rm Pb}$~\cite{frois}
clearly pinpoints the value of the central charge density in this nucleus.
By multiplying this number by $A/Z$ one obtains the relevant central density
of heavy nuclei, corresponding to 0.16 nucleons/${\rm fm}^3$ or $k_F = 1.33~ 
{\rm fm}^{-1}$.
Since the presence of nucleons at the center of a heavy nucleus is confined
to $s$ nucleons, and their depletion is dominated by SRC,
one may conclude that the actual value of the saturation density of
nuclear matter must also be closely linked to the effects
of SRC.
While this argument is particularly appropriate for the deeply bound
$1s_{1/2}$ and $2s_{1/2}$ protons, it continues to hold for the $3s_{1/2}$ 
protons which are depleted predominantly by short-range effects (up to 15\%)
and by at most 10\% due to long-range correlations as discussed above.

The binding energy of nuclei or nuclear matter usually
includes only mean-field contributions to the kinetic energy when 
the calculations
are based on perturbative schemes like the hole-line expansion.
With the presence of high-momentum components in the ground state it becomes
relevant to ask what the real kinetic and potential energy of the system
look like in terms of the single-particle strength distributions.
This theoretical result~\cite{miga,kolt}
has the general form (Koltun sum rule)
\begin{equation}
E_0^A = \bra{\Psi_0^A} \hat{H} \ket{\Psi_0^A}
= \frac{1}{2} \sum_{\alpha\beta} \bra{\alpha} T \ket{\beta} n_{\alpha\beta}
+\frac{1}{2} \sum_\alpha \int_{-\infty}^{\epsilon_F}d\omega\ \omega 
S_h(\alpha, \omega)
\label{eq:be}
\end{equation}
in the case when only two-body interactions are involved.
In this equation, $n_{\alpha\beta}$ is the one-body density matrix element
which can be directly obtained from the single-particle propagator.
A delicate balance exists between the repulsive kinetic energy term
and the attractive contribution of the second term in Eq.~(\ref{eq:be})
which samples the single-particle strength weighted by the energy
parameter $\omega$.
When realistic spectral distributions are used to calculate
these quantities surprising results emerge~\cite{mupodi}.
Such calculations for ${}^{16}{\rm O}$ indicate that the
contribution of the
quasihole states to Eq.~(\ref{eq:be}), corresponding to the $1s_{1/2}$, 
$1p_{3/2}$, and $1p_{1/2}$ orbitals, comprise only 37\% of the total energy
leaving 63\% for the continuum terms that represent the spectral strength
associated with the coupling to low-energy 2h1p states.
These contributions therefore contain the presence of high-momentum
components in the nuclear ground state reflecting
the effect of SRC.
Although these high momenta 
account for only 10\% of the particles in the case of ${}^{16}{\rm O}$, their
contribution to the energy is extremely important.
These results give a first indication of the importance of treating
the dressing of nucleons in finite nuclei in determining the binding
energy per particle.
It is therefore reasonable to conclude that a careful study of short-range
correlations including the full fragmentation of the single-particle
strength is relevant for the calculation of the energy per particle
in finite nuclei.
This has the additional advantage that agreement with data from the
\eep\ reaction can be used to gauge the quality of the
theoretical description in determining the energy per particle.
This argument can be turned inside out by noting that an exact representation
of the spectroscopic strength must lead to the correct energy
per particle according to Eq.~(\ref{eq:be}) in the case of the
dominance of two-body interactions.

Pursuing this argument in the case of nuclear matter, while recalling
that the empirical saturation density is apparently dominated by SRC,
we have calculated nuclear saturation
properties focusing solely on the contribution of SRC.
The experimental results discussed above demand furthermore
that the dressing of nucleons in nuclear matter is then taken into account
in order to be consistent with the extensive collection of 
data from the \eep\ reaction
that have become available in recent years.
The self-consistent calculation of nucleon spectral functions
obtained from the contribution to the nucleon self-energy
of ladder diagrams which include
the propagation of these dressed particles, fulfills this requirement.

It is straightforward to write down the equation that involves
the calculation of the effective interaction in nuclear matter
obtained from the sum of
all ladder diagrams while propagating fully dressed particles.
This result is given in a partial wave representation by the following
equation
\begin{eqnarray}
& {} &\bra{k}\Gamma_{LL'}^{JST}(K,\Omega)\ket{k'} = 
\bra{k}V_{LL'}^{JST}(K,\Omega)\ket{k'} \nonumber \\
& + & \sum_{L''} \int_0^\infty dq\ q^2\
\bra{k}V_{LL''}^{JST}(K,\Omega)\ket{q} g_f^{II}(q;K,\Omega)
\bra{q}\Gamma_{LL''}^{JST}(K,\Omega)\ket{k'} ,
\label{eq:lad}
\end{eqnarray}
where $k,k',$ and $q$ denote relative and $K$ the total momentum
involved in the interaction process.
Discrete quantum numbers correspond to total
spin, $S$, orbital angular momentum, $L,L',L''$, and
the conserved
total angular momentum and isospin, $J$ and $T$, respectively.
The energy $\Omega$ and the total momentum $K$ are conserved and act
as parameters that characterize the effective two-body interaction in the
medium.
The critical ingredient in Eq.~(\ref{eq:lad}) is the noninteracting
propagator $g_f^{II}$ which describes the propagation of the particles
in the medium from interaction to interaction.
For fully dressed particles this propagator is given by
\begin{eqnarray}
g_f^{II}(k_1,k_2;\Omega) & = &
\int_{\epsilon_F}^\infty d\omega_1\ \int_{\epsilon_F}^\infty d\omega_2\
\frac{S_p(k_1,\omega_1) S_p(k_2,\omega_2)}
{\Omega - \omega_1 -\omega_2 +i\eta} \nonumber \\
& - &
\int_{-\infty}^{\epsilon_F} d\omega_1\ \int_{-\infty}^{\epsilon_F} d\omega_2\
\frac{S_h(k_1,\omega_1) S_h(k_2,\omega_2)}
{\Omega - \omega_1 -\omega_2 -i\eta} ,
\label{eq:gtwof}
\end{eqnarray}
where individual momenta $k_1$ and $k_2$ have been used instead
of total and relative momenta as in Eq.~(\ref{eq:lad}).
The dressing of the particles is expressed by the use of particle and
hole spectral functions, $S_p$ and $S_h$, respectively.
The particle spectral function, $S_p$, is defined as a particle
addition probability density in a similar way
as the hole spectral function in Eq.~(\ref{eq:spefu}) for removal.
These spectral functions take into account that the particles propagate
with respect to the correlated ground state incorporating the
presence of high-momentum components in the ground state.
This treatment therefore provides the correlated version of the Pauli
principle and leads to substantial modification with respect to
the Pauli principle effects related to the free Fermi gas.
The corresponding propagator is obtained from Eq.~(\ref{eq:gtwof}) by
replacing the spectral functions by strength distributions characterized
by $\delta$-functions as follows
\begin{eqnarray}
S_p(k,\omega) & = & \theta(k-k_F) \delta(\omega-\epsilon(k)) \nonumber \\
S_h(k,\omega) & = & \theta(k_F-k) \delta(\omega-\epsilon(k)) ,
\label{eq:mf}
\end{eqnarray}
which leads to the Galitski-Feynman propagator including hole-hole
as well as particle-particle propagation of particles characterized
by single-particle energies $\epsilon(k)$.
Discarding the hole-hole propagation then yields the 
Brueckner ladder diagrams with the usual Pauli operator for the free
Fermi gas.
The effective interaction obtained by solving Eq.~(\ref{eq:lad}) using
dressed propagators can be used to construct the self-energy
of the particle. With this self-energy the Dyson equation can be solved
to generate a new incarnation of the dressed propagator.
The process can then be continued by constructing anew the dressed
but noninteracting two-particle propagator according to Eq.~(\ref{eq:gtwof}).
At this stage one can return to the ladder equation and so on until
self-consistency is achieved for the complete Green's function
which is then legitimately called a self-consistent Green's function.

While this scheme is easy to present in equations and words, it is quite
another matter to implement it.
The recent accomplishment of implementing this self-consistency
scheme~\cite{libth} builds upon earlier approximate implementations.
The first nuclear-matter spectral functions were obtained for a semirealistic
interaction by employing mean-field propagators in the ladder
equation~\cite{angels}.
Spectral functions for the Reid interaction were obtained by still employing
mean-field propagators in the ladder equation but with the introduction of
a self-consistent gap in the single-particle spectrum to take into account
the pairing instabilities obtained for a realistic 
interaction~\cite{brian,von93}.
The first solution of the effective interaction using dressed propagators
was obtained by employing a parametrization of the spectral 
functions~\cite{chris}.
The calculations employing dressed propagators in determining
the effective interaction
demonstrate that at normal density one no longer runs into
pairing instabilities on account of the reduced density of states associated
with the reduction of the strength of the quasiparticle pole, $z_{k_F}$,
from 1 in the Fermi gas to 0.7 in the case of dressed propagators.
For two-particle propagation this leads to a reduction factor of
$z_{k_F}^2$ corresponding to about 0.5 that is strong enough to push
even the pairing instability in the ${}^3S_1$-${}^3D_1$ channel
to lower densities~\cite{scat}.

The consequences for the scattering process of
interacting particles in nuclear matter characterized
by phase shifts and cross sections are also substantial and lead to a
reduction of the cross section in a wide range of energies~\cite{scat}.
\begin{figure}[htb]
  \begin{center}
    \parbox[b]{0.5\linewidth}{
      \includegraphics[height=.25\textheight]{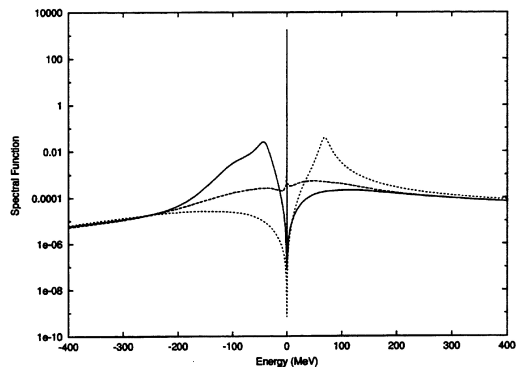}}
    \parbox[b]{7mm}{~}
    \parbox[b]{0.4\linewidth}{
    \caption[hits in strips plus germanium]{\label{fig:spfu}
      Self-consistent spectral functions at $k_F = 1.36$ ${\rm fm}^{-1}$.
      Single-particle momenta corresponding to $k = 0$ (solid), $k_F$ (dashed),
      and 2.1 ${\rm fm}^{-1}$ (dotted) are shown.
   }}
  \end{center}
\end{figure}
The current implementation of the self-consistent scheme for the
propagator across the summation of all ladder diagrams
includes a parametrization of the imaginary part of the nucleon self-energy.
Employing a representation in terms of two gaussians above and two below
the Fermi energy, it is possible to accurately represent the
nucleon self-energy as generated by the contribution of relative $S$-waves
(and including the tensor coupling to the ${}^3D_1$ channel)~\cite{libth}.
Self-consistency at a density corresponding to $k_F = 1.36\ {\rm fm}^{-1}$
is achieved in about ten iteration steps, each involving a considerable
amount of computer time~\cite{libth}.
\begin{figure}[htb]
  \begin{center}
    \parbox[b]{0.5\linewidth}{
      \includegraphics[height=.2\textheight]{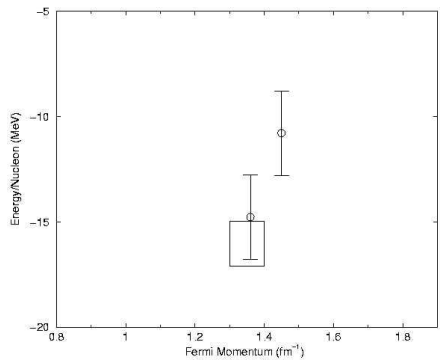}}
    \parbox[b]{7mm}{~}
    \parbox[b]{0.4\linewidth}{
    \caption[hits in strips plus germanium]{\label{fig:bea2}
      The energy per particle calculated at two densities. The saturation 
      density for this self-consistent Green's function calculation with the
      Reid potential is possibly in agreement with the empirical result.}}
  \end{center}
\end{figure}
It is important to reiterate that this
scheme isolates the contribution of short-range correlations
to the energy per particle which is obtained from Eq.~(\ref{eq:be}).
An important result pertaining to this ``second generation'' spectral functions
is shown in Fig.~\ref{fig:spfu} related to the emergence of a common tail
at large negative energy for different momenta. Such a common 
tail was previously obtained at high energy~\cite{von91} as a signature of SRC.
This common tail may play a significant role in generating some additional
binding energy at lower densities.
At present, results for two densities corresponding to $k_F = 1.36$ and 1.45
${\rm fm}^{-1}$ have been obtained.
Self-consistency is achieved for the contribution of the ${}^1S_0$ and 
${}^3S_1$-${}^3D_1$ channels to the self-energy. The other partial wave 
contributions have been added separately. The corresponding results for the
binding energy have been obtained by averaging the parametrizations of the
corresponding self-energies. The difference between these results
also provides us with a conservative estimate
of the lack of self-consistency including higher partial waves.
This error estimate is included in Fig.~\ref{fig:bea2} for the energy per
particle calculated from the energy (Koltun) sum rule in Eq.~(\ref{eq:be}). 
These results suggest that it is possible to obtain reasonable saturation
properties for nuclear matter provided one only includes SRC in the 
determination of the equation of state.
Clearly, this assertion implies that long-range contributions
to the energy per particle need not be considered in explaining nuclear
saturation properties.
Considerations relevant to this issue are presented elsewhere~\cite{wim}.

\section{Conclusions}
A review of experimental data that exhibit clear evidence for the notion
that nucleons in nuclei are dressed particles is given.
A recent doubt concerning the validity of the interpretation of the
spectroscopic strength has been resolved by showing that the same spectroscopic
factors can be used to explain data from the \eep\ reaction at different values
of $Q^2$.
Based on these considerations and the success of theoretical calculations
to account for the qualitative features of the single-particle
strength distributions, it is suggested that the dressing of nucleons
must be taken into account in calculations of the energy per particle.
By identifying the dominant contribution of SRC
to the empirical saturation density, it is argued that these correlations
need to be isolated in the study of nuclear matter.
A scheme which fulfills this requirement and includes the propagation
of dressed particles, as required by experiment, is outlined.
Successful implementation of this scheme has recently been
demonstrated~\cite{libth} for the continuum version.
A discrete version has been implemented by the Gent group~\cite{yves,gent}.
These new calculations may lead to new insight into
the long-standing problem of nuclear saturation.

\section*{Acknowledgments}
This work was supported by the U. S. National Science Foundation under Grant
No. PHY-9900713.

\end{document}